\documentclass[10pt,prd,singlecolumn,
nofootinbib,
noshowkeys,
noshowpacs,
superscriptaddress,
floatfix
]{revtex4-2}
\usepackage{amsmath,amsfonts,amsthm,amssymb}
\usepackage[dvips]{graphics,graphicx}
\usepackage[usenames,dvipsnames]{xcolor}
\definecolor{darkblue}{RGB}{0,0,196}
\definecolor{darkgreen}{RGB}{0,120,0}
\usepackage[colorlinks=true,linktocpage=true,linkcolor=darkblue,citecolor=red,urlcolor=darkblue]{hyperref}
\usepackage{cancel}
\usepackage{bbold}
\usepackage{multirow}
\usepackage{longtable}
\usepackage{color}
\usepackage[normalem]{ulem}
\usepackage{hyperref}
\usepackage{bigints}
\usepackage{xparse}
\usepackage{physics}
\usepackage{verbatim}
\usepackage{minibox}
\usepackage{comment}
\usepackage{appendix}
\usepackage{slashed}
\usepackage{marginnote}
\usepackage{graphicx}
\usepackage[nice]{nicefrac}
\usepackage{amsmath}
\usepackage[colorinlistoftodos]{todonotes}
\usepackage{hepunits}
\usepackage{stackengine,scalerel}
\newcommand\hstar[1]{\ThisStyle{\ensurestackMath{%
\setbox0=\hbox{$\SavedStyle#1$}%
\stackengine{0pt}{\copy0}{\kern.2\ht0\smash{\SavedStyle\star}}{O}{c}{F}{T}{S}}}}
\definecolor {darkgreen}{rgb}{0.2,0.7,0.2}

\begin{document}
\title{Stability and causality of rest frame modes in second-order spin hydrodynamics }
\author{Asaad Daher}
\email{asaad.daher@ifj.edu.pl}
\affiliation{Institute  of  Nuclear  Physics  Polish  Academy  of  Sciences,  PL-31-342  Krak\'ow,  Poland}
\author{Wojciech Florkowski}
\email{wojciech.florkowski@uj.edu.pl}
\affiliation{Institute of Theoretical Physics, Jagiellonian University, PL-30-348 Krak\'ow, Poland}
\author{Radoslaw Ryblewski}
\email{radoslaw.ryblewski@ifj.edu.pl}
\affiliation{Institute  of  Nuclear  Physics  Polish  Academy  of  Sciences,  PL-31-342  Krak\'ow,  Poland}
\author{Farid Taghinavaz}
\email{ftaghinavaz@ipm.ir}
\affiliation{School of Particles and Accelerator, Institute for Research in Fundamental Sciences (IPM), P. O. Box 19395-5531, Tehran, Iran.}
%
%*****************************************************************
%
\begin{abstract}
We analyze the low- and high-momentum rest frame modes in the second-order spin hydrodynamics and check the asymptotic causality of the theory. A truncation scheme of the Israel-Stewart formalism derived in our earlier work is proposed that extends the minimal causal formulation. It consists of altogether 40 interconnected relaxation-type dynamical equations -- 16 (24) of them correspond to the independent components of the energy-momentum (spin) tensor. Similarly to previous studies, we find that the stability of the perturbations and asymptotic causality require using the spin equation of state that satisfies the generalized Frenkel condition demanding that the ``electric'' and ``magnetic'' components of the spin density tensor have opposite signs. For low-momentum modes this behavior is similar to that found earlier for the first-order (Navier-Stokes) spin hydrodynamics.  
\end{abstract}
\maketitle
%**************************************
%
\section{Introduction}
Investigations of the stability and causality properties of relativistic spin hydrodynamics have attracted a lot of attention lately~\cite{Hattori:2019lfp,Daher:2022wzf, Sarwar:2022yzs, Xie:2023gbo, Weickgenannt:2023btk,Daher:2024ixz}. They shed light on the mathematical structure of the formalism and may determine its usefulness for description of realistic systems, for example, those produced in heavy-ion collisions~\cite{STAR:2017ckg, STAR:2018gyt,ALICE:2019onw,ALICE:2019aid,STAR:2020xbm,Kornas:2020qzi,STAR:2021beb,Niida:2018hfw,STAR:2019erd,ALICE:2021pzu,Becattini:2024uha,Niida:2024ntm}. They can also be useful to distinguish between different formulations of spin hydrodynamics, the most popular of which are: the spin conserving approach \cite{Florkowski:2017ruc,Florkowski:2017dyn,Florkowski:2018ahw,Florkowski:2018fap,Bhadury:2020puc,Bhadury:2020cop}, the gradient expansion (of the first and second order~\cite{Hattori:2019lfp,Fukushima:2020ucl,Daher:2022wzf,Daher:2022xon,Biswas:2022bht,Biswas:2023qsw}) combined with the condition of positive entropy production, kinetic theory with non-local interactions~\cite{Weickgenannt:2022zxs,Speranza:2020ilk,Wagner:2022amr,Sheng:2021kfc,Weickgenannt:2023nge,Weickgenannt:2020aaf,Weickgenannt:2019dks,Wagner:2023cct,Weickgenannt:2022qvh,Wagner:2022gza}, and quantum statistical mechanics~\cite{Becattini:2011zz,Becattini:2012pp,Becattini:2007nd,Becattini:2009wh,Becattini:2018duy,Becattini:2014yxa,Becattini:2023ouz}, see also~\cite{She:2021lhe,Gallegos:2021bzp,Hongo:2021ona,Li:2020eon,Shi:2020htn,Peng:2021ago,Hu:2021lnx,Abbasi:2022rum,Bemfica:2023res,Shokri:2023rpp,Gavassino:2023qnw,Almaalol:2022pjc,Rocha:2023ilf,Elfner:2022iae,Strickland:2024moq,An:2023yfq,Ambrus:2022qya,Obukhov:2023gyr,Ambrus:2022yzz,Ambrus:2021sjg,Obukhov:2023yti,Weickgenannt:2023bss}. Stability and causality analyses based on the gradient expansion use (in the leading order) the phenomenological form of the spin tensor~\cite{Weyssenhoff:1947iua}
\begin{equation}
S^{\lambda, \mu\nu} = u^\lambda S^{\mu\nu},
\label{eq:spintensor}
\end{equation}
where $u^\lambda$ is the hydrodynamic flow and $S^{\mu\nu}$ is the {\it spin density tensor}. Recently, a connection between the phenomenological form~\eqref{eq:spintensor} and the canonical formalism, where the spin tensor is totally antisymmetric, has been established~\cite{Daher:2022xon}. Moreover, the spin density~\eqref{eq:spintensor} is taken in the form 
\begin{equation}
S^{\mu \nu} = S(T,\mu) \omega^{\mu \nu}, 
 \label{eq:S}
\end{equation}
where $\omega^{\mu \nu}$ is the {\it spin polarization tensor} (namely, the spin chemical potential $\Omega^{\mu\nu}$ divided by the temperature $T$), while $S(T,\mu)$ is a scalar function of temperature and chemical potential. We note that the relation (\ref{eq:S}) can be treated as a~kind of constraint imposed on the spin equation of state. 

In this work we generalize the results obtained in Refs.~\cite{Hattori:2019lfp,Daher:2022wzf, Sarwar:2022yzs,Xie:2023gbo}. In Refs.~\cite{Hattori:2019lfp,Daher:2022wzf, Sarwar:2022yzs} the first-order formulation was used, while in Ref.~\cite{Xie:2023gbo} the minimal causal spin hydrodynamics was studied. The latter is defined as an analog of the minimal causal extension of conventional hydrodynamics. It concentrates on the essential terms in the second order of gradient expansion to get a causal theory. Our present analysis can be treated as a generalization of the minimal causal spin hydrodynamics, as we consider the full 40 relaxation-type dynamical equations derived in~\cite{Biswas:2023qsw} (with 16 (24) equations corresponding to the independent components of the energy-momentum (spin) tensor). The only truncation we propose is to neglect the so-called mixed terms. We systematically repeat the linear stability analysis of the rest frame low- and high-momentum modes. Dispersion relations for all possible excitations are derived. We find that the stability conditions for the extended model are essentially the same as those found before in the first-order and minimally causal theories. Hence, going to higher orders and including more terms in the second-order theory is not crucial for the stability of the perturbations. Furthermore, we check the asymptotic causality of our framework and find that causality and stability conditions agree. 

As the matter of fact, the fundamental property responsible for the stability is a specific dependence of the spin density tensor on the components of the spin polarization tensor, namely, the derivatives of the spin density tensor with respect to the ``electric'' ($\omega^{0i}$) and ``magnetic'' ($\omega^{ij}$) components of the polarization tensor in the fluid rest frame should have opposite signs. This means that the relation $S^{\mu \nu} = S(T,\mu) \omega^{\mu \nu}$ should be replaced by a more general formula, for example, by the expression~\cite{Daher:2024ixz}
\begin{equation}
S^{\gamma \delta} = (S_1-S_2) \left(\omega^{\gamma \alpha} u_\alpha u^\delta -  \omega^{\delta \alpha} u_\alpha u^\gamma \right) + S_2 \omega^{\gamma \delta},
\label{eq:newS}
\end{equation}
where $S_1$ and $S_2$ are two different functions of $T$ and $\mu$. In the fluid rest frame, where $u^\mu = (1,0,0,0)$, one finds
\begin{equation}
 S^{0i} = S_1 \omega^{0i}, \quad  S^{ij} = S_2 \omega^{ij}. 
 \label{eq:S0iSij}
\end{equation}
The stability conditions found in earlier works are
\begin{equation}
\chi_b \equiv \frac{\partial S^{0i}}{\partial \omega^{0i} } < 0, \qquad
\chi_s \equiv \frac{\partial S^{ij}}{\partial \omega^{ij} } > 0,
\label{eq:mainresult}
\end{equation}
where the derivatives are taken in the reference frame where the considered system rests as a whole and is unpolarized. This leads to the conditions $S_1 < 0$ and $S_2 > 0$.  We note that the use of Eq.~(\ref{eq:newS}) instead of Eq.~(\ref{eq:S}) may be treated as a minimal improvement to achieve stability. In general, more complex structures for the spin tensor may be considered (see, for example, Refs~\cite{Florkowski:2017ruc,Florkowski:2017dyn,Florkowski:2018ahw,Florkowski:2018fap,Bhadury:2020puc,Bhadury:2020cop}).

It is important to emphasize that the frameworks discussed herein also assume the following thermodynamic relations~\footnote{Research employing a quantum statistical approach has recently demonstrated that these relations  may generally require modifications, as outlined in~\cite{Becattini:2023ouz}.}
\begin{eqnarray}
\varepsilon + p = T s   + S^{\alpha \beta} \omega_{\alpha \beta}, \qquad
dp = s dT + S^{\alpha \beta} d\omega_{\alpha \beta}.
\label{eq:therm}
\end{eqnarray}
Equations (\ref{eq:therm}) combined with the conservation laws and the condition of positive entropy production uniquely determine the structure of spin hydrodynamic equations without invoking an explicit form of the spin density tensor. Hence, we can use the results obtained before and decide to choose between (\ref{eq:S}) or (\ref{eq:newS}) only at the last stage of the stability analysis. 

Altogether, our analysis demonstrates that both the first- and second-order theories exhibit stability of the rest frame low-momentum modes provided the proper spin equation of state is used. This finding agrees with conventional relativistic hydrodynamics~\cite{Hiscock:1985zz} and other relevant studies~\cite{Kovtun:2019hdm, Bemfica:2019knx, Bemfica:2017wps, Koide:2006ef, Denicol:2008ha, Romatschke:2009im, Pu:2009fj}. Moreover, the same conditions provide the high momentum stability and asymptotic causality of the considered framework (provided the used relaxation times are sufficiently large). Similar theoretical explorations employing linear mode analysis have also been conducted in the context of relativistic magneto-hydrodynamics~\cite{Fang:2024skm} and chiral hydrodynamics~\cite{Speranza:2021bxf,Abboud:2023hos}. While we focus here on rest frame studies, our future work could delve into the stability, causality, and possible correlations between the two in a boosted frame. One novel approach leverages the information current introduced recently in Refs.~\cite{Gavassino:2021kjm,Gavassino:2022isg}, especially considering its recent application in minimal spin hydrodynamics~\cite{Ren:2024pur}.

This work is organized as follows: Sec.~\ref{sec2} introduces our "40=16+24" relaxation-time dynamical equations, which lays the foundation for the subsequent linear mode analysis in Sec.~\ref{sec3}.  Sec.~\ref{section4} investigates the stability of the system in both low- and high-momentum regimes. Section~\ref{sec5} focuses on the asymptotic causality analysis. Finally, Sec.~\ref{sec6} presents our conclusions and outlines potential areas for future research.

\bigskip
Throughout the text we use the metric tensor  $g_{\mu\nu}= \hbox{diag}(+, -, -, -)$. The projector orthogonal to $u^{\mu}$ is defined as $\Delta^{\mu\nu}\equiv g^{\mu\nu}-u^{\mu}u^{\nu}$. The partial derivative operator can be decomposed into two parts, one along the flow direction and the other orthogonal to it, i.e., $\partial_{\mu}=u_{\mu}D+\nabla_{\mu}$ where $D\equiv u^{\mu}\partial_{\mu}$ and $\nabla_{\mu}\equiv\Delta_{\mu}^{~\alpha}\partial_{\alpha}$. The expansion rate is defined as $\theta\equiv \partial_{\mu}u^{\mu}$. For symmetric and antisymmetric part of arbitrary tensor $X^{\mu\nu}$ we use the notation $X^{\mu\nu}_{(s)}\equiv X^{(\mu\nu)}=(X^{\mu\nu}+X^{\nu\mu})/2$ and $X^{\mu\nu}_{(a)}\equiv X^{[\mu\nu]}=(X^{\mu\nu}-X^{\nu\mu})/2$, respectively. Projection orthogonal to $u^{\mu}$ of a four-vector $X^\mu$ is represented as $X^{\langle\mu\rangle}\equiv \Delta^{\mu\nu}X_{\nu}$. A symmetric, orthogonal and traceless part of $X^{\mu\nu}$ is denoted as $X^{\langle\mu\nu\rangle}\equiv \Delta^{\mu\nu}_{\alpha\beta}X^{\alpha\beta}\equiv \frac{1}{2}\left(\Delta^{\mu}_{~\alpha}\Delta^{\nu}_{~\beta}+\Delta^{\mu}_{~\beta}\Delta^{\nu}_{~\alpha}-\frac{2}{3}\Delta^{\mu\nu}\Delta_{\alpha\beta}\right)X^{\alpha\beta}$. Similarly, $X^{\langle[\mu\nu]\rangle}\equiv \Delta^{[\mu\nu]}_{[\alpha\beta]}X^{\alpha\beta}\equiv \frac{1}{2}\left(\Delta^{\mu}_{~\alpha}\Delta^{\nu}_{~\beta}-\Delta^{\mu}_{~\beta}\Delta^{\nu}_{~\alpha}\right)X^{\alpha\beta}$ denotes the antisymmetric and orthogonal projection. 
%
%*************************************
%
\section{Israel-Stewart-like Evolution Equation}
\label{sec2}
The framework of relativistic spin hydrodynamics is based on the conservation laws for energy, momentum, and angular momentum. They can be represented by the following differential equations
\begin{align}
    &\partial_{\mu}T^{\mu\nu}=0,\label{EMTconservation}\\
    & \partial_{\mu}J^{\mu\alpha\beta} =\partial_{\mu}S^{\mu\alpha\beta} +2 T^{[\alpha\beta]}=0\label{TAMconservation}.
\end{align}
Here $T^{\mu\nu}$ is the energy-momentum tensor (EMT), while $J^{\mu\alpha\beta}=L^{\mu\alpha\beta}+S^{\mu\alpha\beta}$ is the total angular momentum tensor, with $L^{\mu\alpha\beta}=2 x^{[\alpha}T^{\mu\beta]}$ being its orbital part and $S^{\mu\alpha\beta}=-S^{\mu\beta\alpha}$ its spin part. We use the following decompositions
\begin{align}
&T^{\mu\nu}=T^{\mu\nu}_{0}+T^{\mu\nu}_{1s}+T^{\mu\nu}_{1a}=(\varepsilon+p)u^{\mu}u^{\nu}-pg^{\mu\nu}
+2h^{(\mu}u^{\nu)}+\pi^{\mu\nu}+\Pi\Delta^{\mu\nu}+2q^{[\mu}u^{\nu]}
+\phi^{\mu\nu},~\label{EMT}\\
&S^{\mu\alpha\beta}=S^{\mu\alpha\beta}_{0}+S^{\mu\alpha\beta}_{1}=u^{\mu} S^{\alpha\beta}+2u^{[\alpha}\Delta^{\mu\beta]}\Phi+2u^{[\alpha}\tau^{\mu\beta]}_{s}+2u^{[\alpha}\tau^{\mu\beta]}_{a}+\Theta^{\mu\alpha\beta},~\label{Spintensor}
\end{align}
where $T^{\mu\nu}_{0}$ is the equilibrium (perfect fluid) energy-momentum tensor, $T^{\mu\nu}_{1s}$ and $T^{\mu\nu}_{1a}$ are the symmetric and antisymmetric EMT dissipative corrections, whereas  $S^{\mu\alpha\beta}_{0}$ and $S^{\mu\alpha\beta}_{1}$ represent the zeroth- and first-order contributions to the spin tensor~\cite{Biswas:2023qsw}.

The energy-momentum tensor $T^{\mu\nu}$ can typically have 16 independent components corresponding to: the energy density $\varepsilon$, fluid 4-velocity $u^{\mu}$, heat flux $h^{\mu}$ and its antisymmetric analog $q^{\mu}$, the shear-stress tensor $\pi^{\mu\nu}$ and its antisymmetric counterpart $\phi^{\mu\nu}$, and the bulk pressure $\Pi$. The variables $u^{\mu}$, $h^{\mu}$ and $q^{\mu}$ each have 3 independent components due to the conditions $u^{\mu}u_{\mu}=1$, $h^{\mu}u_{\mu}=0$ and $q^{\mu}u_{\mu}=0$. Both $\pi^{\mu\nu}$ and $\phi^{\mu\nu}$ are orthogonal to $u^{\mu}$. Furthermore, $\pi^{\mu\nu}$ is symmetric and traceless, while $\phi^{\mu\nu}$ is antisymmetric. Hence, $\pi^{\mu\nu}$ has 5 independent components, and $\phi^{\mu\nu}$ has 3. The bulk pressure $\Pi$ is a scalar. This counting gives 19 degrees of freedom rather than 16. Therefore, we adopt the Landau frame with $h^{\mu}=0$. The spin tensor $S^{\mu\alpha\beta}$ has in total 24 independent components, where the spin density $S^{\alpha\beta}=u_{\mu}S^{\mu\alpha\beta}$ has 6 components and is considered to be of the leading order in the gradient expansion, i.e., $S^{\alpha\beta}\sim\mathcal{O}(1)$. The quantity $\Phi$ is a scalar, $\tau_{s}^{\mu\beta}$ is symmetric traceless and orthogonal to $u_{\mu}$ so it has 5 components, while $\tau_{a}^{\mu\beta}$ is antisymmetric and transverse to the $u_{\mu}$ so it has 3 components. Finally, $\Theta^{\mu\alpha\beta}$ is antisymmetric orthogonal to $u_{\mu}$ in all indices, hence it has only 9 components.

Studies examining stability and causality properties of the relativistic second-order relativistic hydrodynamics without spin~\cite{Pu:2009fj, Betz:2008me, Song:2008si, Song:2008si} were done by focusing on the Naiver-Stokes terms and those proportional to the relaxation times. The resulting equations are known in the literature as "simplified I-S equations" and have simple consequences -- dissipative quantities such as $\pi^{\mu\nu}, \Pi,...$ relax to their corresponding Navier-Stokes values on time scales determined by the appropriate relaxation times $\tau_{\pi^{\mu\nu}}, \tau_{\Pi},...$ . Recently, a study of the stability and causality properties for the second-order spin hydrodynamics has been performed following similar strategy as outlined above~\cite{Xie:2023gbo}. In this work, we extend this approach by considering 40=16+24 relaxation-type dynamical equations (where 16 corresponds to the independent components of the energy-momentum tensor and 24 to those of the spin tensor) without neglecting all the second-order contributions. Therefore, we will keep non-mixing terms sourcing from $Q^{\mu}$ and those due to $\partial_{\mu}\left(\beta\omega_{\alpha\beta}\right)S^{\mu\alpha\beta}_{1}$ appearing at the level of entropy-current (see Eq.~(21) in Ref.~\cite{Biswas:2023qsw} for details). Technically, this is equivalent to using
\begin{align}
    \partial_{\mu}s^{\mu}_{\rm IS}=&T^{\mu\nu}_{1a}\left(\partial_{\mu}\beta_{\nu}+2\beta\omega_{\mu\nu}\right)+T^{\mu\nu}_{1s}\partial_{\mu}\beta_{\nu}+\partial_{\mu}Q^{\mu}-\partial_{\mu}\left(\beta\omega_{\alpha\beta}\right)S^{\mu\alpha\beta}_1,
\end{align}
where
\begin{align}
Q^{\mu} = & ~~u^{\mu}\left(a_{1}\Pi^2+a_{2}\pi^{\lambda\nu}\pi_{\lambda\nu}+a_{4}q^{\lambda}q_{\lambda}+a_{5}\phi^{\lambda\nu}\phi_{\lambda\nu}\right)\nonumber
\\
& + u^{\mu}\left(\tilde{a}_{1}\Phi^{2}+\Tilde{a}_{2}\tau_{s}^{\lambda\nu}\tau_{s\lambda\nu}+\Tilde{a}_{3}\tau_{a}^{\lambda\nu}\tau_{a\lambda\nu}+\Tilde{a}_{4}\Theta^{\lambda\alpha\beta}\Theta_{\lambda\alpha\beta}\right).
\end{align}
Note that $Q^{\mu}$ neglects mixing terms proportional to $\Pi q^{\mu}, \pi^{\mu\nu}q_{\nu}, \Theta^{\alpha\beta\mu}\phi_{\alpha\beta}$,... see Ref.~\cite{Biswas:2023qsw}. Here $\beta_{\mu}=u_{\mu}/T$ is the thermal velocity, $\omega_{\mu\nu}$ is the spin polarization tensor ($\omega_{\mu\nu}\sim\mathcal{O}(\partial)$ in our hydrodynamic counting scheme), while $a_{i}'s$ and $\tilde{a}_{i}'s$ are dimension-full coefficients. Therefore, the evolution equations for the energy, momentum, and dissipative parts of the energy-momentum tensor derived in~\cite{Biswas:2023qsw} are reduced to:  
\begin{align}
D\varepsilon+(\varepsilon+p)\theta=\pi^{\mu\nu}\partial_{\mu}u_{\nu}+\Pi\theta-\nabla\cdot q+\phi^{\mu\nu}\partial_{\mu}u_{\nu}\,,\label{energyeq}
\end{align}
\begin{align}
(\varepsilon+p) Du^{\alpha} - \nabla^{\alpha}  p=
&-\Delta^{\alpha}_{\nu}\partial_{\mu}\pi^{\mu\nu}-\Delta^{\mu\alpha}\partial_{\mu}\pi+\pi D u^{\alpha}-q^{\mu}\partial_{\mu}u^{\alpha}\nonumber\\
&+\Delta^{\alpha}_{\nu}Dq^{\nu}+q^{\alpha}\theta-\Delta^{\alpha}_{\nu}\partial_{\mu}\phi^{\mu\nu}\, ,
\end{align}
\begin{align}
    &\tau_{\Pi}D\Pi+\Pi=\zeta\left[\theta+Ta_{1}\Pi\theta+T\Pi Da_{1}\right],
\end{align}
\begin{align}
\tau_{\pi}\Delta^{\mu\nu}_{\alpha\beta}D\pi^{\alpha\beta}+\pi^{\mu\nu}=2\eta\left[(\nabla^{(\mu}u^{\nu)}-\frac{1}{3}\theta\Delta^{\mu\nu})+Ta_{2}\theta\pi^{\mu\nu}+T\pi^{\mu\nu}Da_{2}\right],
\end{align}
\begin{align}
\tau_{q}\Delta^{\mu}_{\nu}Dq^{\nu}+q^{\mu}=\lambda\left[(\beta\nabla^{\mu}T+Du^{\mu}-4\omega^{\mu\nu}u_{\nu})-Ta_{4}q^{\mu}\theta-Tq^{\mu}Da_{4}\right],\label{qeq}
\end{align}
\begin{align}
\tau_{\phi}\Delta^{[\mu\nu]}_{[\alpha\beta]}D\phi^{\alpha\beta}+\phi^{\mu\nu}=\gamma\left[ (\beta\nabla^{[\mu}u^{\nu]}+2\beta\Delta^{\mu\alpha}\Delta^{\nu\beta}\omega_{\alpha\beta})+a_{5}\theta\phi^{\mu\nu}+\phi^{\mu\nu}Da_{5}\right]\label{phieq},
\end{align}
where $\zeta\geq 0,\eta\geq 0,\lambda\geq 0$, and $\gamma\geq 0$ are the corresponding transport coefficients. The relaxation times of various dissipative quantities are defined as, $\tau_{\Pi}=-2a_1\zeta T\geq 0$, $\tau_{\pi}=-4a_2\eta T\geq 0$, $\tau_{q}=2a_4\lambda T\geq 0$, and $\tau_{\phi}=-2a_5\gamma\geq 0$~\cite{Biswas:2023qsw}. Similarly, the evolution equations of the spin density and spin dissipative currents reduce to: 
\begin{align}
DS^{\alpha\beta}+S^{\alpha\beta}\theta+\partial_{\mu}S^{\mu\alpha\beta}_{1}=-2(q^{\alpha}u^{\beta}-q^{\beta}u^{\alpha}+\phi^{\alpha\beta}),
\end{align}
\begin{align}
    \tau_{\Phi}D\Phi+\Phi=\chi_{1}\left[-2u^{\alpha}\nabla^{\beta}(\beta\omega_{\alpha\beta})+\tilde{a}_{1}\theta\Phi+\Phi D\tilde{a}_{1}\right],\label{Phieq}
\end{align}
\begin{align}
\tau_{\tau_{s}}\Delta^{\mu\nu}_{\alpha\beta}D\tau^{\alpha\beta}_{s}+\tau^{\mu\nu}_{s}=\chi_{2}\bigg[&-u^{\alpha}(\Delta^{\gamma\mu}\Delta^{\rho\nu}+\Delta^{\gamma\nu}\Delta^{\rho\mu}-\frac{2}{3}\Delta^{\gamma\rho}\Delta^{\mu\nu})\nabla_{\gamma}(\beta\omega_{\alpha\rho})+\tilde{a}_{2}\theta\tau^{\mu\nu}_{s}+\tau^{\mu\nu}_{s}D\tilde{a}_{2}\bigg],
\label{tauseq}
\end{align}
\begin{align}
\tau_{\tau_{a}}\Delta^{[\mu\nu]}_{[\alpha\beta]}D\tau_{a}^{\alpha\beta}+\tau^{\mu\nu}_{a}=\chi_{3}\left[-u^{\alpha}(\Delta^{\gamma\mu}\Delta^{\rho\nu}-\Delta^{\gamma\nu}\Delta^{\rho\mu})\nabla_{\gamma}(\beta\omega_{\alpha\rho})+\tilde{a}_{3}\theta\tau_{a}^{\mu\nu}+\tau^{\mu\nu}_{a}D\tilde{a}_{3}\right],
\label{tauaeq}
\end{align}
\begin{align}  \tau_{\Theta}\Delta^{\alpha}_{\lambda}\Delta^{\mu}_{\sigma}\Delta^{\nu}_{\beta}D\Theta^{\lambda\sigma\beta}+\Theta^{\alpha\mu\nu}=-\chi_{4}\left[-\Delta^{\delta\mu}\Delta^{\rho\nu}\Delta^{\gamma\alpha}\nabla_{\gamma}(\beta\omega_{\delta\rho})+\tilde{a}_{4}\theta\Theta^{\alpha\mu\nu}+\Theta^{\alpha\mu\nu}D\tilde{a}_{4}\right],
\label{Thetaeq}
\end{align}
where $\chi_{1}\geq 0,\chi_{2}\geq 0,\chi_{3}\geq 0,$ and $\chi_{4}\geq 0$ are the new spin transport coefficients. Various spin-relaxation times can be identified as, $\tau_{\Phi}=-2\tilde{a_1}\chi_1\geq 0$, $\tau_{\tau_s}= -2\tilde{a}_2\chi_2\geq 0$, $\tau_{\tau_a}= -2\tilde{a}_3\chi_3\geq 0$, and $\tau_{\Theta}= 2\tilde{a}_4\chi_4\geq 0$~\cite{Biswas:2023qsw}. As previously discussed, our formalism is composed of 40=16+24 equations corresponding to 40 unknowns, therefore the system is mathematically closed. 
%
%*******************************************************
\section{Linear mode analysis}
\label{sec3}
We investigate the dynamic evolution of hydrodynamic and non-hydrodynamic modes within a spinful relativistic fluid. This is achieved through the examination of linear perturbations superimposed on a state of global equilibrium and solving the hydrodynamic equations. Our focus lies on perturbations characterized by the following form:
\begin{align}
&\varepsilon(x)\rightarrow\varepsilon_{0}+\delta\varepsilon(x),~~~~~~~~~~u^{\mu}\rightarrow u^{\mu}_{0}+\delta u^{\mu}(x)=(1,\Vec{0})+(0,\delta \vec{v}),\nonumber\\
&\omega^{\mu\nu}(x)\rightarrow 0 +\delta \omega^{\mu\nu}(x),~~~~S^{\mu\nu}(x)=0+\delta S^{\mu\nu}(x),\nonumber\\
&X(x)\rightarrow 0+\delta X(x).\label{perturbations}
\end{align}
where $X$ represents the behavior under linear perturbation of various dissipative currents of the energy-momentum~\eqref{EMT} and spin tensors~\eqref{Spintensor}. Below, we introduce the equation(s) of state and the constants as follows:
\begin{align}
\label{definition(s)}
&\delta p=c_{s}^{2}\delta \varepsilon,~~\delta T=\frac{T_{0}\,c_{s}^{2}}{\varepsilon_{0}+p_{0}}\delta\varepsilon,~~\chi_{b}=\frac{\partial S^{i0}}{\partial\omega^{i0}}<0,~~\chi_{s}=\frac{\partial S^{ij}}{\partial\omega^{ij}}>0,\nonumber\\
&D_{b}=\frac{4\lambda}{\chi_{b}},~~D_{s}=\frac{4\gamma}{\chi_{s}},~~\lambda^{'}=\frac{\lambda}{\varepsilon_{0}+p_{0}},~~\gamma^{'}=\frac{\gamma\beta_{0}}{2},\nonumber\\
&\tilde{\chi}_{1}=\frac{2\beta_{0}}{\chi_{b}}\chi_{1},~~{\tilde{\chi}_{2}=\frac{\beta_{0}}{\chi_{b}}\chi_{2}},~~{\tilde{\chi}_{3}=\frac{\beta_{0}}{\chi_{b}}\chi_{3}},~~{\tilde{\chi}_{4}=\frac{\beta_{0}}{\chi_{s}}\chi_{4}},
\end{align}
where $c_{s}^{2}$ is the speed of sound. By linearizing the spin hydrodynamic equations (Eqs.~\eqref{energyeq}--\eqref{Thetaeq}) with respect to the perturbation specified in Eq.~\eqref{perturbations}, we arrive at: 
\begin{align}
\partial_{0}\delta\varepsilon+(\varepsilon_{0}+p_{0})\partial_{i}\delta u^{i}+\partial_{i}\delta q^{i}=0,\label{energyperturbation}
\end{align}
\begin{align}
    (\varepsilon_{0}+p_{0})\partial_{0}\delta u^{i}-c_{s}^{2}\partial^{i}\delta\varepsilon-\partial_{0}\delta q^{i}+\partial_{\mu}\delta \pi^{\mu i}+\partial^{i}\delta\pi+\partial_{\mu}\delta\phi^{\mu i}=0,
\end{align}
\begin{align}
    \tau_{\Pi}\partial_{0}\delta\Pi+\delta\Pi-\zeta\partial_{i}\delta u^{i}=0\, ,
\end{align}
\begin{align}
    \tau_{\pi}\partial_{0}\delta\pi^{ij}+\delta \pi^{ij}-\eta(\partial^{i}\delta u^{j}+\partial^{j}\delta u^{i})+\frac{2}{3}\eta\Delta^{ij}_{0}\partial_{k}\delta u^{k}=0\, ,
\end{align}
\begin{align}
    \tau_{q}\partial_{0}\delta q^{i}+\delta q^{i}-\lambda^{'}c_{s}^{2}\partial^{i}\delta\varepsilon-\lambda\partial_{0}\delta u^{i}+D_{b}\delta S^{i0}=0 ,
\end{align}
\begin{align}
   \tau_{\phi}\partial_{0}\delta \phi^{ij}+\delta \phi^{ij}-\gamma^{'}(\partial^{i}\delta u^{j}-\partial^{j}\delta u^{i})-\frac{\beta_{0}}{2}D_{s}\delta S^{ij}=0, \,
\end{align}
\begin{align}
    \partial_{0}\delta S^{0i}+\Delta^{ki}_{0}\partial_{k}\delta\Phi+\partial_{k}\delta\tau_{s}^{ki}+\partial_{k}\delta\tau^{ki}_{a}-2\delta q^{i}=0\, ,
\end{align}
\begin{align}
    \partial_{0}\delta S^{ij}+\partial_{k}\delta\Theta^{kij}+2\delta\phi^{ij}=0\, ,
\end{align}
\begin{align}
    \tau_{\Phi}\partial_{0}\delta\Phi
+\delta\Phi-{\tilde{\chi}_{1}\partial_{i}\delta S^{i0}}=0\, ,
\end{align}
\begin{align}
\tau_{\tau_{s}}\partial_{0}\delta\tau_{s}^{ij}+\delta \tau_{s}^{ij}{-\tilde{\chi}_{2}(\partial^{i}\delta S^{j0}+\partial^{j}\delta S^{i0})+\frac{2}{3}\tilde{\chi}_{2}\Delta^{ij}_{0}\partial_{k}\delta S^{k0}}=0\, ,
\end{align}
\begin{align}
    \tau_{\tau_{a}}\partial_{0}\delta\tau_{a}^{ij}+\delta \tau_{a}^{ij}+{\tilde{\chi}_{3}(-\partial^{i}\delta S^{j0}+\partial^{j}\delta S^{i0})}=0\, ,
\end{align}
\begin{align}
\tau_{\Theta}\partial_{0}\delta\Theta^{ijk}+\delta\Theta^{ijk}-{\tilde{\chi}_{4}\partial^{i}\delta S^{jk}}=0.\label{Thetaperturbation}
\end{align}
The above perturbations $\delta\varepsilon$, $\delta\Pi$, $\delta S^{ij}$, $\delta \Theta^{ijk}$,... etc., can be expressed as plane waves: 
\begin{align}
& \delta\varepsilon=\widetilde{\delta\varepsilon}~e^{-i\omega t+i\vec{k}\cdot\vec{x}},~~~\delta\Pi=\widetilde{\delta\Pi}~e^{-i\omega t+i\vec{k}\cdot\vec{x}},\nonumber\\
& \delta S^{ij}=\widetilde{\delta S}^{ij}~e^{-i\omega t+i\vec{k}\cdot\vec{x}},~~~\delta \Theta^{ijk}=\widetilde{\delta \Theta}^{ijk}~e^{-i\omega t+i\vec{k}\cdot\vec{x}}~\label{waveequation}.
\end{align}
Since the system exhibits rotational symmetry, we can focus on waves propagating only in the $z$ direction, with a~wave vector $\vec{k} = (0, 0, k_z)$. For such a choice, Eqs.~\eqref{energyperturbation}--\eqref{Thetaperturbation} can be expressed as a matrix equation, where a $40\times40$ block-diagonal matrix $\mathbf{M}_{40\times40}$ multiplies a vector $\mathbf{V}$ composed of the 40 Fourier components of the fluctuations, namely, we derive the condition $\mathbf{M}_{40\times40} \, \mathbf{V}=0$, where
\begin{align}
&\mathbf{M}_{40\times40}=\left(
\begin{array}{ccccc}
\mathbf{A}_{10\times10} & 0 & 0 & 0 & 0 \\
0 & \mathbf{B}_{9\times9} & 0 & 0 & 0 \\
0 & 0 & \mathbf{B}_{9\times9} & 0 & 0 \\
0 & 0 & 0 & \mathbf{C}_{3\times3}& 0\\
0 & 0 & 0 & 0 & \mathbf{D}_{9\times9}
\end{array}
\right), \qquad
\mathbf{V}=(\mathbf{v}_{\mathbf{A}},\mathbf{v}_{\mathbf{B}_{\mathbf{x}}},\mathbf{v}_{\mathbf{B}_{\mathbf{y}}},\mathbf{v}_{\mathbf{C}},\mathbf{v}_{\mathbf{D}})^{\intercal}
\end{align}
and
\begin{align}
&\mathbf{v}_{\mathbf{A}}=\left(\delta \tilde{\varepsilon},\delta \tilde{u}^{z},\delta \tilde{\pi}^{xx},\delta\tilde{\pi}^{yy},\delta \tilde{\Pi},\delta \tilde{q}^{z},\delta \tilde{\Phi},\delta\tilde{S}^{0z},\delta \tilde{\tau}^{xx}_{s},\delta \tilde{\tau}^{yy}_{s}\right),\nonumber\\
&\mathbf{v}_{\mathbf{B_{x}}}=\left(\delta \tilde{u}^{x},\delta \tilde{q}^{x},\delta\tilde{\pi}^{zx},\delta\tilde{\phi}^{zx},\delta \tilde{S}^{0x},\delta \tilde{\tau}^{zx}_{s},\delta \tilde{\tau}_{a}^{zx},\delta \tilde{S}^{xz},\delta \tilde{\Theta}^{zxz}\right),\nonumber\\
&\mathbf{v}_{\mathbf{B}_{\mathbf{y}}}=\left(\delta \tilde{u}^{y},\delta \tilde{q}^{y},\delta\tilde{\pi}^{zy},\delta\tilde{\phi}^{zy},\delta \tilde{S}^{0y},\delta \tilde{\tau}^{zy}_{s},\delta \tilde{\tau}_{a}^{zy},\delta \tilde{S}^{yz},\delta \tilde{\Theta}^{zyz}\right),\nonumber\\
&\mathbf{v}_{\mathbf{C}}=(\delta \tilde{S}^{xy},\delta \tilde{\phi}^{xy},\delta \tilde{\Theta}^{zxy}),\nonumber\\
&\mathbf{v}_{\mathbf{D}}=\left(\delta \tilde{\pi}^{xy},\delta \tilde{\tau}^{xy}_{s},\delta \tilde{\tau}_{a}^{xy},\delta\tilde{\Theta}^{xxy},\delta\tilde{\Theta}^{xxz},\delta\tilde{\Theta}^{yyz},\delta\tilde{\Theta}^{xyz},\delta\tilde{\Theta}^{yxz},\delta\tilde{\Theta}^{yxy}\right).
\label{fluctuations}
\end{align}
The block parts of the matrix $\mathbf{M}_{40\times40}$ represent 3 groups of coupled channels and one group of decoupled channels. The coupled channels are: the sound channels described by the matrix $\mathbf{A}_{10\times10}$, the shear channels described by $\mathbf{B}_{9\times9}$ (appearing twice), and a group of purely spin channels defined by $\mathbf{C}_{3\times3}$ that is distinctive for spin hydrodynamics. The decoupled channels are defined by $\mathbf{D}_{9\times9}$. The explicit forms for the matrices $\mathbf{A}_{10\times10}$, $\mathbf{B}_{9\times9}$, $\mathbf{C}_{3\times3}$, and $\mathbf{D}_{9\times9}$ are:

\begin{eqnarray}
\mathbf{A}_{10\times10}=\left(
\begin{array}{cccccccccc}
 -i \omega  & i k_{z} & 0 & i (\varepsilon_{0}+p_{0})k_{z} & 0 & 0 & 0 & 0 & 0 & 0 \\
 i c_{s}^{2} k_{z} & i \omega  & 0 & -i (\varepsilon_{0}+p_{0}) \omega  & -i k_{z} & -i k_{z} & -i k_{z} & 0 & 0 & 0 \\
 0 & 0 & 0 & -i k_{z} \zeta  & 1-i \tau_{\Pi} \omega  & 0 & 0 & 0 & 0 & 0 \\
 0 & 0 & 0 & \frac{-2}{3}i \eta k_{z} & 0 & 1-i\tau_{\pi} \omega  & 0 & 0 & 0 & 0 \\
 0 & 0 & 0 & -\frac{2}{3}i \eta k_{z} & 0 & 0 & 1-i \tau_{\pi} \omega  & 0 & 0 & 0 \\
 i c_{s}^{2}\lambda^{'} k_{z} & 1-i\tau_{q} \omega  & -D_{b} & i \lambda  \omega  & 0 & 0 & 0 & 0 & 0 & 0 \\
 0 & -2 & -i \omega  & 0 & 0 & 0 & 0 & -i k_{z} & -i k_{z} & -i k_{z} \\
 0 & 0 & i \tilde{\chi}_{1} k_{z} & 0 & 0 & 0 & 0 & 0 & 0 & 1-i \tau_{\Phi} \omega  \\
 0 & 0 & \frac{2}{3} i \tilde{\chi}_{2} k_{z} & 0 & 0 & 0 & 0 & 1-i\tau_{\tau_{s}} \omega  & 0 & 0 \\
 0 & 0 & \frac{2}{3} i \tilde{\chi_{2}} k_{z} & 0 & 0 & 0 & 0 & 0 & 1-i \tau_{\tau_{s}} \omega  & 0 \\
\end{array}
\right), \nonumber \\
\end{eqnarray}

\begin{eqnarray}
\mathbf{B}_{9\times9}=\left(
\begin{array}{ccccccccc}
 i \omega  & 0 & 0 & -i(\varepsilon_{0}+p_{0}) \omega  & 0 & i k_{z} & 0 & 0 & i k_{z} \\
 0 & 0 & 0 & i \eta k_{z} & 0 & 1-i\tau_{\pi} \omega  & 0 & 0 & 0 \\
 1-i \tau_{q} \omega  & -D_{b} & 0 & i \lambda  \omega  & 0 & 0 & 0 & 0 & 0 \\
 0 & 0 & -\frac{D_{s}}{2T_{0}} & -i \gamma^{'}k_{z} & 0 & 0 & 0 & 0 & -1+i\tau_{\phi} \omega  \\
 -2 & -i \omega  & 0 & 0 & 0 & 0 & ik_{z} & i k_{z} & 0 \\
 0 & 0 & -i \omega  & 0 & ik_{z} & 0 & 0 & 0 & -2 \\
 0 & -i \tilde{\chi}_{2} k_{z} & 0 & 0 & 0 & 0 & 0 & 1-i \tau_{\tau_{s}} \omega  & 0 \\
 0 & -i \tilde{\chi}_{3} k_{z} & 0 & 0 & 0 & 0 & 1-i \tau_{\tau_{a}} \omega  & 0 & 0 \\
 0 & 0 & i \tilde{\chi}_{4}k_{z} & 0 & 1-i \tau_{\Theta} \omega  & 0 & 0 & 0 & 0 \\
\end{array}
\right), \nonumber \\
\end{eqnarray}

\begin{eqnarray}
\mathbf{C}_{3\times3}=\left(
\begin{array}{ccc}
 -\frac{D_{s}}{2T_{0}} & 0 & 1-i \tau_{\phi} \omega  \\
 -i \omega  & i k_{z} & 2 \\
 i \tilde{\chi}_{4}k_{z} & 1-i\tau_{\Theta} \omega  & 0 \\
\end{array}
\right), \nonumber \\
\end{eqnarray}

\begin{eqnarray}
\mathbf{D}_{9\times9}=\left(
\begin{array}{ccccccccc}
 1-i\tau_{\tau_{s}} \omega  & 0 & 0 & 0 & 0 & 0 & 0 & 0 & 0 \\
 0 & 1-i \tau_{\tau_{a}} \omega  & 0 & 0 & 0 & 0 & 0 & 0 & 0 \\
 0 & 0 & 1-i \tau_{\Theta} \omega  & 0 & 0 & 0 & 0 & 0 & 0 \\
 0 & 0 & 0 & 1-i \tau_{\Theta} \omega  & 0 & 0 & 0 & 0 & 0 \\
 0 & 0 & 0 & 0 & 1-i \tau_{\Theta} \omega  & 0 & 0 & 0 & 0 \\
 0 & 0 & 0 & 0 & 0 & 1-i \tau_{\Theta} \omega   & 0 & 0 & 0 \\
 0 & 0 & 0 & 0 & 0 & 0 & 1-i \tau_{\Theta} \omega  & 0 & 0 \\
 0 & 0 & 0 & 0 & 0 & 0 & 0 & 1-i \tau_{\Theta} \omega  & 0 \\
 0 & 0 & 0 & 0 & 0 & 0 & 0 & 0 & 1-i\tau_{\pi} \omega  \\
\end{array}
\right). \nonumber \\
\end{eqnarray}
Due to the block-diagonal form of the matrix $\mathbf{M}$, \footnote{In places where we do not emphasize the importance of the dimension of a considered matrix, we skip the index containing this information. We also use the symbols from $\mathbf{A}$ to $\mathbf{D}$ to denote the corresponding groups of channels. } its determinant becomes the product 
\begin{align}
\textrm{det}(\mathbf{M})&=\textrm{det}(\mathbf{A})\,\textrm{det}(\mathbf{B})^{2}\,\textrm{det}(\mathbf{C})\,\textrm{det}(\mathbf{D}).\label{determinant}
\end{align}
The dispersion relations are determined by the solutions of the equation $\textrm{det}(\mathbf{M})=0$. In relativistic hydrodynamics without spin, the linearized equilibrium fluctuations are commonly split into the "sound channel" and "shear channel" modes~\cite{Kovtun:2019hdm}. Below, we perform a similar classification for the case including spin. 
%
%%%%%%%%%%%%%%%%%%%%%%%%%%%%%%%%%%%%%%%%%%%%%%%%%%%%%%
%
\subsection{Low-momentum Modes }
\label{sec3.1}
\subsubsection{Coupled sound channels \texorpdfstring{$\mathbf{A}$}{Lg}}
The sound channel describes longitudinal fluctuations. They are parallel to the flow direction $(0,0,k_z)$. We may think of them as sound waves propagating alongside the moving fluid. In our case, the sound channels can be identified with the fluctuations forming the vector $\mathbf{v}_{\mathbf{A}}$ in Eq.~\eqref{fluctuations}. By analyzing the determinant of the matrix $\mathbf{A}$ in Eq.~\eqref{determinant}, we can isolate the frequencies associated with such longitudinal waves, and find the following expressions:
\begin{align}
    &\omega_{1,2}=\pm c_{s}k_{z}-i\frac{(\frac{4}{3}\eta+\zeta)}{2(\varepsilon_{0}+p_{0})}k_{z}^{2},\label{oso12}\\
    &\omega_{3}=-\frac{i}{\tau_{\Phi}}-i\frac{\tilde{\chi}_{1}}{\Sigma_{\Phi}}k_{z}^{2},\\
    &\omega_{4}=-\frac{i}{\tau_{\pi}}+\frac{4}{3}i\frac{\eta}{(\varepsilon_{0}+p_{0})\Lambda_{\pi}}k_{z}^{2},\\
    &\omega_{5}=-\frac{i}{\tau_{\Pi}}+i\frac{\zeta}{(\varepsilon_{0}+p_{0})\Lambda_{\Pi}}k_{z}^{2},\\
    &\omega_{6}=-\frac{i}{\tau_{\tau_{s}}}-\frac{4}{3}i\frac{\tilde{\chi}_{2}}{\Sigma_{\tau_{s}}}k_{z}^{2},\\
    &\omega_{7,8}=-i\frac{\left(1\pm \sqrt{8D_{b}(\tau_{q}-\lambda^{'})+1}\right)}{2(\tau_{q}-\lambda^{'})}+\mathcal{O}(k_{z}^{2}).\label{oso11}
\end{align}
In Eqs.~\eqref{oso12}--\eqref{oso11}, we have introduced the notation
\begin{align}
&\Sigma_{\Phi}=1+\frac{2D_{b}\tau_{\Phi}^{2}}{(\tau_{\Phi}-\tau_{q})+\lambda^{'}},~~\Sigma_{\tau_{s}}=1+\frac{2D_{b}\tau_{\tau_{s}}^{2}}{\left(\tau_{\tau_{s}}-\tau_{q}\right)+\lambda^{'}},\nonumber\\
&\Lambda_{\pi}=1+\frac{\lambda^{'}}{2D_{b}\tau_{\pi}^{2}+\tau_{\pi}-\tau_{q}},~~\Lambda_{\Pi}=1+\frac{~\lambda^{'}}{2D_{b}\tau_{\Pi}^{2}+\tau_{\Pi}-\tau_{q}}.
\end{align}
We note that it is expected to have here 10 dispersion relations as the matrix $\mathbf{A}$ is 10-dimensional. We display above only eight of them, as two purely damped modes (for small values of $k_z$) are shifted down to the group $\mathbf{D}$ (that becomes 11-dimensional in this way). In the following subsection, we are going to discuss the shear channels.
%
%%%%%%%%%%%%%%%%%%%%%%%%%%%%%%%%%%%%%%%%%%%%%%%%%%%%%%
%
\subsubsection{Degenerate coupled shear channels \texorpdfstring{$\mathbf{B}$}{Lg}}
The shear channels encompass fluctuations that deviate from the main flow direction, essentially being orthogonal to the momentum direction $(0,0,k_z)$. They can be imagined as waves rippling perpendicularly to the overall flow of matter. In our specific case, the shear channels can be identified with the fluctuations forming the vectors $\mathbf{v}_{\mathbf{B}_{\mathbf{x}}}$ and $\mathbf{v}_{\mathbf{B}{\mathbf{y}}}$ (as defined by Eq.~\eqref{fluctuations}). These vectors represent displacements in the $x$ and $y$ directions, respectively. Similarly to the previous case, the shear channel modes are obtained from the condition $\textrm{det}(\mathbf{B})^{2} = 0$, where the square implies that each shear mode in the group $\mathbf{B}$ is doubly degenerate. After lengthy but straightforward calculations we find
\begin{align}
    &\omega_{1}=-i\frac{\eta}{(\varepsilon_{0}+p_{0})}k_{z}^{2},\label{osh1}\\
    &\omega_{2}=-\frac{i}{\tau_{\pi}}+i\frac{\eta}{(\varepsilon_{0}+p_{0})\Lambda_{\pi}}k_{z}^{2},\\
    &\omega_{3}=-\frac{i}{\tau_{\tau_{\Theta}}}+i\frac{\tilde{\chi}_{4}}{\Upsilon_{\Theta}}k_{z}^{2},\\
    &\omega_{4}=-\frac{i}{\tau_{\tau_{a}}}-i\frac{\tilde{\chi}_{3}}{\Sigma_{\tau_{a}}}k_{z}^{2},\\
    &\omega_{5}=-\frac{i}{\tau_{\tau_{s}}}-i\frac{\tilde{\chi}_{2}}{\Sigma_{\tau_{s}}}k_{z}^{2},\\%~~\Sigma_{\tau_{s}}=1+\frac{2D_{b}\tau_{\tau_{s}}^{2}}{\left(\tau_{\tau_{s}}-\tau_{q}\right)+\lambda^{'}}\\
    &\omega_{6,7}=-i\frac{\left(1\pm\sqrt{1-4D_{s}\tau_{\phi}\beta_{0}}\right)}{2\tau_{\phi}}+\mathcal{O}(k_{z}^{2}),\\
    &\omega_{8,9}=-i\frac{\left(1\pm \sqrt{8D_{b}(\tau_{q}-\lambda^{'})+1}\right)}{2(\tau_{q}-\lambda^{'})}+\mathcal{O}(k_{z}^{2}).\label{osh89}
\end{align}
Here we have introduced the notation
\begin{align}
    \Sigma_{\tau_{a}}=1+\frac{2D_{b}\tau_{\tau_{a}}^{2}}{\left(\tau_{\tau_{a}}-\tau_{q}\right)+\lambda^{'}},~~\Upsilon_{\Theta}=1+D_{s}\beta_{0}\left(\frac{\tau_{\Theta}^{2}}{\tau_{\phi}-\tau_{\Theta}}\right).
\end{align}
%
%%%%%%%%%%%%%%%%%%%%%%%%%%%%%%%%%%%%%%%%%%%%%%%%%%%%%%
%
\subsubsection{Coupled channels \texorpdfstring{$\mathbf{C}$}{Lg}}
Our exploration of relativistic hydrodynamics with spin indicates existence of new phenomena going beyond the familiar sound and shear fluctuations. They correspond to channels $\mathbf{C}$ and are given by the vector $\mathbf{v}_{\mathbf{C}}$ defined by Eq.~\eqref{fluctuations}. Studying the determinant of the matrix $\mathbf{C}$ in Eq.~\eqref{determinant}, we obtain:
\begin{align}
    &\omega_{1}=-\frac{i}{\tau_{\Theta}}+i\frac{\tilde{\chi}_{4}}{\Upsilon_{\Theta}}k_{z}^{2},\\
    &\omega_{2}=-i\frac{\left(1-\sqrt{1-4D_{s}\tau_{\phi}\beta_{0}}\,\right)}{2\tau_{\phi}}-i\frac{\tilde{\chi}_{4}\tau_{\phi}\left(1+\sqrt{1-4D_{s}\tau_{\phi}\beta_{0}}\right)}{-2\left(2D_{s}\tau_{\Theta}\beta_{0}-\sqrt{1-4D_{s}\tau_{\phi}\beta_{0}}\right)\tau_{\phi}+\left(1-\sqrt{1-4D_{s}\tau_{\phi}\beta_{0}}\right)\tau_{\Theta}}k_{z}^{2},\\
    &\omega_{3}=-i\frac{\left(1+\sqrt{1-4D_{s}\tau_{\phi}\beta_{0}}\,\right)}{2\tau_{\phi}}-i\frac{\tilde{\chi}_{4}\tau_{\phi}\left(-1+\sqrt{1-4D_{s}\tau_{\phi}\beta_{0}}\right)}{2\left(2D_{s}\tau_{\Theta}\beta_{0}+\sqrt{1-4D_{s}\tau_{\phi}\beta_{0}}\right)\tau_{\phi}-\left(1+\sqrt{1-4D_{s}\tau_{\phi}\beta_{0}}\right)\tau_{\Theta}}k_{z}^{2}.
\end{align}
The fluctuations included in $\mathbf{C}$ involve spin densities and antisymmetric part of the energy-momentum tensor. Investigating their physical properties requires separate research.
%
%%%%%%%%%%%%%%%%%%%%%%%%%%%%%%%%%%%%%%%%%%%%%%%%%%%%%%%%%%%%%%%%%%%%
%
\subsubsection{Decoupled non-propagating channels \texorpdfstring{$\mathbf{D}$}{Lg}}
Finally, we present the purely damped modes included in the group $\mathbf{D}$. They correspond to the deviations described by the vector $\mathbf{v}_{\mathbf{D}}$ defined in~\eqref{fluctuations}. The frequencies are obtained again by solving the equation $\textrm{det}\mathbf{(D)} = 0$. In this way we find:
\begin{align}
    &\omega_{1,2}=-\frac{i}{\tau_{\pi}},\\
    &\omega_{3\rightarrow 8}=-\frac{i}{\tau_{\Theta}},\\
    &\omega_{9}=-\frac{i}{\tau_{\tau_{a}}},\\
    &\omega_{10,11}=-\frac{i}{\tau_{\tau_{s}}}.
\end{align}
%
%%%%%%%%%%%%%%%%%%%%%%%%%%%%%%%%%%%%%%
%
\subsection{High-momentum Modes}
We now turn on to the characteristics of the modes within the high-momentum limit. It's crucial to emphasize that the channel classification framework we have outlined in Section~\ref{sec3.1} remains entirely valid in the high-momentum regime. From a physical standpoint, when the wavenumber along the z-direction, $k_z$, assumes large values, it becomes particularly insightful to prioritize our analysis on two categories of terms: those constituting the leading order and those exhibiting proportionality to the momentum itself. This allows us to extract the most essential information about the modes.
%
%%%%%%%%%%%%%%%%%%%%%%%%%%%%%%%%%%%%%%%%%%%%%%%%%%%%%%%%%%%%%%%%%%%%%%%
%
\subsubsection{Coupled sound channels \texorpdfstring{$\mathbf{A}$}{Lg}}
Similar to the low-momentum case, the longitudinal fluctuations, vibrations parallel to the flow direction, are the focus of the sound channel. The sound channels can be identified with the fluctuations represented by the vector $\mathbf{v}_{\mathbf{A}}$ in Eq.~\eqref{fluctuations}. By analyzing the determinant of matrix $\mathbf{A}$ in Eq.~\eqref{determinant} in the high-momentum limit we obtain the following modes,
\begin{align}\label{eq:AC-HM}
    &{\omega_1 = -\frac{i ( \frac{4}{3} \eta + \xi)}{\frac{4}{3} \eta \tau_\Pi+ \xi \tau_\pi} %-\frac{4 i (\varepsilon_0 + p_0) \eta \xi (\tau_\pi - \tau_\Pi)^2 \left(\frac{4}{3} \eta (3\lambda' + \tau_q - \tau_\Pi) + \xi (3\lambda' + \tau_q - \tau_\pi)\right)}{ 3 \lambda' (\frac{4}{3} \eta \tau_\Pi+ \xi \tau_\pi)^4 k_z^2}
    + \mathcal{O}(\frac{1}{k_z^2}) + \mathcal{O}(\frac{1}{k_z^4}),}\\
&{\omega_2 = - \frac{i(\frac{4}{3}\tilde{\chi}_2 + \tilde{\chi}_1)}{\frac{4}{3}\tilde{\chi}_2 \tau_\Phi + \tilde{\chi}_1 \tau_{\tau_s}} %+ \frac{4 i (\tau_{\tau_s} - \tau_\Phi)^2 \tilde{\chi}_1 \tilde{\chi}_2 (\frac{4}{3}\tilde{\chi}_2 + \tilde{\chi}_1)}{3 k_z^2 (\frac{4}{3}\tilde{\chi}_2 \tau_\Phi + \tilde{\chi}_1 \tau_{\tau_s})^4}
    +\mathcal{O}(\frac{1}{k_z^2}) + \mathcal{O}(\frac{1}{k_z^4}),}\\
    &{\omega_{3,4} =-\frac{i}{2 \tau_{\tau_s} \tau_\Phi} \frac{  \tau_{\tau_s}^2 \tilde{\chi}_1 + \frac{4}{3} \tau_\Phi ^2 \tilde{\chi}_2}{ \tau_{\tau_s} \tilde{\chi}_1 + \frac{4}{3} \tau_\Phi \tilde{\chi}_2} \pm \sqrt{-\frac{\frac{4}{3}\tilde{\chi}_2 \tau_\Phi + \tilde{\chi}_1 \tau_{\tau_s}}{\tau_{\tau_s} \tau_\Phi}}\, k_z + \mathcal{O}(\frac{1}{k_z})},\\
    &{\omega_{s_{1},s_{2}} = -i \frac{\sum\limits_{n=0}^3 c_n v_{s_1,s_2}^{2n}}{\sum\limits_{n=0}^3 d_n v_{s_1,s_2}^{2n}} + v_{s_1,s_2} \, k_z + \mathcal{O}(\frac{1}{k_z})}.
\end{align}
{For the sake of simplicity, here, $\omega_{s_{1},s_{2}}$ represents 4 modes with $s_1 = \pm$ and $s_2 = \pm$. The 4 modes can be thought of as $\omega_{+,+},\,\omega_{+,-},\,\omega_{-,+},\,\omega_{-,-}$, where}
\begin{align}
    &{v_{s_1,s_2} = s_1 \sqrt{\frac{\mathcal{A} + s_2 \mathcal{B}}{2 (\varepsilon_0 + p_0) (\tau_q - \lambda') \tau_\pi \tau_\Pi}}},\nonumber\\
    &{\mathcal{A} \equiv c_s^2 (\varepsilon_0 + p_0) (\tau_q + 3 \lambda') \tau_\pi \tau_\Pi + \tau_q (\frac{4}{3} \eta \tau_\Pi+ \xi \tau_\pi)},\nonumber\\
&{\mathcal{B}^2 \equiv \mathcal{A}^2 - 4 c_s^2 \tau_\pi \tau_\Pi \lambda  (\tau_q - \lambda') (\frac{4}{3} \eta \tau_\Pi+ \xi \tau_\pi)},\nonumber
\end{align}
{along with the terms,}
\begin{align}
    &{c_0 = -c_s^2 \lambda' \bigg(\tilde{\chi}_1 \left(\frac{4\eta}{3} (\tau_\Pi + \tau_{\tau_s}) + \xi (\tau_\pi + \tau_{\tau_s})\right) + \frac{4 \tilde{\chi}_2}{3} \left(\frac{4\eta}{3} (\tau_\Pi + \tau_\Phi) + \xi (\tau_\pi + \tau_\Phi)\right)\bigg),} \nonumber\\
    &{c_1= \frac{1}{9} \bigg(-3 c_s^2 \bigg(\lambda' \bigg(4 \eta  (\tau_\Pi  \tau_\Phi +\tau_{\tau_s} (\tau_\Pi +\tau_\Phi ))+3 \xi  (\tau_\pi  \tau_\Phi +\tau_{\tau_s} (\tau_\pi +\tau_\Phi ))}\nonumber\\
    &{\qquad-3 (\varepsilon_0 + p_0)\left(\tau_\pi  \tau_\Pi(  3 \tilde{\chi}_1+4  \tilde{\chi}_2)+4 \tau_\Phi   \tilde{\chi}_2 (\tau_\pi  + \tau_\Pi ) +3 \tau_{\tau_s} \tilde{\chi}_1 (\tau_\pi +\tau_\Pi )\right)\bigg)}\nonumber\\
    &{-(\varepsilon_0 + p_0) \bigg(\tau_q \left(\tau_\pi  \tau_\Pi(  3 \tilde{\chi}_1+4  \tilde{\chi}_2)+4 \tau_\Phi   \tilde{\chi}_2 (\tau_\pi  + \tau_\Pi ) +3 \tau_{\tau_s} \tilde{\chi}_1 (\tau_\pi +\tau_\Pi )\right)+\tau_\pi  \tau_\Pi  (3 \tau_{\tau_s} \tilde{\chi}_1+4 \tau_\Phi  \tilde{\chi}_2)\bigg)\bigg)} \nonumber\\
    &{+4 \eta  \bigg(\tau_q ( \tau_\Pi(  3 \tilde{\chi}_1+4  \tilde{\chi}_2)+3 \tau_{\tau_s} \tilde{\chi}_1+4 \tau_\Phi  \tilde{\chi}_2)+\tau_\Pi  (3 \tau_{\tau_s} \tilde{\chi}_1+4 \tau_\Phi  \tilde{\chi}_2)\bigg)}\nonumber\\
  & {+3 \xi  \bigg(\tau_q ( \tau_\pi  (3 \tilde{\chi}_1+4  \tilde{\chi}_2)+3 \tau_{\tau_s} \tilde{\chi}_1+4 \tau_\Phi  \tilde{\chi}_2)+\tau_\pi  (3 \tau_{\tau_s} \tilde{\chi}_1+4 \tau_\Phi  \tilde{\chi}_2)\bigg)\bigg),}\nonumber
  \end{align}
    \begin{align}
    &{c_2= \frac{1}{3} \bigg((\varepsilon_0 + p_0) \bigg(3 c_s^2 \bigg(3 \lambda' (\tau_\pi  \tau_\Pi  \tau_\Phi +\tau_\pi  \tau_{\tau_s} (\tau_\Pi +\tau_\Phi )+\tau_\Pi  \tau_{\tau_s} \tau_\Phi )+\tau_\pi  \tau_\Pi  \tau_q (\tau_\Phi + \tau_{\tau_s}) + \tau_{\tau_s} \tau_\Phi \left(\tau_q  (\tau_\pi  +\tau_\Pi)   + \tau_\pi  \tau_\Pi  \right)\bigg)}\nonumber\\
    &{+\lambda' \bigg(\tau_\pi  (3 \tau_\Pi  \tilde{\chi}_1+4 \tau_\Pi  \tilde{\chi}_2+4 \tau_\Phi  \tilde{\chi}_2)+4 \tau_\Pi  \tau_\Phi  \tilde{\chi}_2+3 \tau_{\tau_s} \tilde{\chi}_1 (\tau_\pi +\tau_\Pi )\bigg)}\nonumber\\
    &{-\tau_\pi  \tau_\Pi (4  \tau_\Phi  \tilde{\chi}_2 + 3   \tau_q \tilde{\chi}_1) - 4 \tau_q \tilde{\chi}_2 \left(\tau_\pi(\tau_\Pi + \tau_\Phi)+\tau_\Pi   \tau_\Phi  \right)
    -3 \tau_{\tau_s} \tilde{\chi}_1 \left(\tau_q (\tau_\pi   +\tau_\Pi ) +  \tau_\pi  \tau_\Pi  \right)\bigg)}\nonumber\\
    &{+4 \eta  \left(\tau_\Pi  \tau_q \tau_\Phi +\tau_q \tau_{\tau_s} (\tau_\Pi +\tau_\Phi )+\tau_\Pi  \tau_{\tau_s} \tau_\Phi \right)+3 \xi  \left(\tau_\pi  \tau_q \tau_\Phi +\tau_q \tau_{\tau_s} (\tau_\pi +\tau_\Phi )+\tau_\pi  \tau_{\tau_s} \tau_\Phi \right)\bigg),}\nonumber\\
 &{c_3 = -(\varepsilon_0 + p_0) \bigg(-\lambda' \left(\tau_\pi  \tau_\Pi  \tau_\Phi +\tau_{\tau_s} \tau_\Phi  (\tau_\pi +\tau_\Pi )+\tau_\pi  \tau_\Pi  \tau_{\tau_s}\right) + \tau_\Phi  \left(\tau_\Pi  \tau_q (\tau_\pi +\tau_{\tau_s})+\tau_\pi  \tau_q \tau_{\tau_s}+\tau_\pi  \tau_\Pi  \tau_{\tau_s}\right)+ \tau_\pi  \tau_\Pi  \tau_q \tau_{\tau_s}\bigg),}\nonumber
    \end{align}
{and,} 
\begin{align}
&{d_0 = -2c_s^2 \lambda' \left(\frac{4}{3} \eta \tau_\Pi+ \xi \tau_\pi\right) \left(\frac{4}{3}\tilde{\chi}_2 \tau_\Phi + \tilde{\chi}_1 \tau_{\tau_s}\right),}\nonumber\\ %\quad d_3 = 8 (\varepsilon_0 + p_0) (\lambda' - \tau_q) \tau_{\tau_s} \tau_\pi \tau_\Pi \tau_\Phi,\nonumber\\
    &{d_1 = 4 \bigg(\tau_q \left(\frac{4}{3} \eta \tau_\Pi+ \xi \tau_\pi\right) \left(\frac{4}{3}\tilde{\chi}_2 \tau_\Phi + \tilde{\chi}_1 \tau_{\tau_s}\right) + c_s^2 \left(-\lambda' \tau_{\tau_s} \tau_\Phi \left(\frac{4}{3} \eta \tau_\Pi+ \xi \tau_\pi\right)  + (\varepsilon_0 + p_0) \tau_\pi \tau_\Pi(3\lambda' + \tau_q) \left(\frac{4}{3}\tilde{\chi}_2 \tau_\Phi + \tilde{\chi}_1 \tau_{\tau_s}\right)\right)\bigg),}\nonumber\\
    &{d_2 = 6 \tau_q \tau_{\tau_s} \tau_\Phi \left(\frac{4}{3} \eta \tau_\Pi+ \xi \tau_\pi\right) + 6 (\varepsilon_0 + p_0) \tau_\pi \tau_\Pi \left((\lambda' - \tau_q) \left(\frac{4}{3}\tilde{\chi}_2 \tau_\Phi + \tilde{\chi}_1 \tau_{\tau_s}\right) + c_s^2 \tau_{\tau_s} \tau_\Phi (3\lambda' +\tau_q)\right),}\nonumber\\
    &{d_3 = 8 (\varepsilon_0 + p_0) (\lambda' - \tau_q) \tau_{\tau_s} \tau_\pi \tau_\Pi \tau_\Phi.}\nonumber
\end{align}
%
%%%%%%%%%%%%%%%%%%%%%%%%%%%%%%%%%%%%%%%%%%%%%%%%%%%%%%%%%%%%%%%%%
%
\subsubsection{Degenerate coupled shear channels \texorpdfstring{$\mathbf{B}$}{Lg}}
{Shear channels deal with deviations perpendicular to the main flow, essentially orthogonal to the momentum direction $(0,0,k_z)$. In similar process, yet in the high-momentum limit, as in~\eqref{sec3.1} we obtain}
\begin{align}\label{eq:BC-HM}
    &{\omega_{1} = -\frac{i}{\tau_q} %-\frac{i}{k_z^2 \tau_q^4} \left(\frac{\lambda (\tau_q -\tau_\pi)(\tau_q -\tau_\phi)}{\gamma' (\tau_\pi -\tau_q) + \eta (\tau_\phi -\tau_q)} + \frac{2 D_b \tau_q^2 (\tau_{\tau_a} -\tau_q)(\tau_q - \tau_{\tau_s})}{\tau_{\tau_a} \tilde{\chi}_2 + \tau_{\tau_s} \tilde{\chi}_3 - \tau_q (\tilde{\chi}_2 + \tilde{\chi}_3)} \right)
    + \mathcal{O}(\frac{1}{k_z^2})
    + \mathcal{O}(\frac{1}{k_z^4}),}\\
    &{\omega_{2} = -\frac{i (\gamma' + \eta)}{\gamma' \tau_\pi + \eta \tau_\phi}% -\frac{i \eta  \left(\frac{D_s (\gamma' \tau_\pi +\eta  \tau_\phi )^2 (\gamma' (\tau_\pi -\tau_\Theta )+\eta  (\tau_\phi -\tau_\Theta ))}{T_0 \tilde{\chi}_4}+\frac{ (\varepsilon_0 + p_0)\gamma' (\gamma'+\eta ) (\tau_\pi -\tau_\phi )^2 (\gamma' (\lambda'+\tau_\pi -\tau_q)+\eta  (\lambda'-\tau_q+\tau_\phi ))}{\gamma' (\tau_\pi -\tau_q)+\eta  (\tau_\phi -\tau_q)}\right)}{k_z^2 (\gamma' \tau_\pi +\eta  \tau_\phi )^4}
    + \mathcal{O}(\frac{1}{k_z^2})+ \mathcal{O}(\frac{1}{k_z^4}),}\\
    &{\omega_3 = - \frac{i (\tilde{\chi}_2 + \tilde{\chi}_3)}{\tau_{\tau_a}  \tilde{\chi}_2 + \tau_{\tau_s}\tilde{\chi}_3} %+ \frac{i \tilde{\chi}_2 \tilde{\chi}_3 (\tau_{\tau_a}-\tau_{\tau_s})^2 \left(2 D_b \tau_{\tau_a}^2 \tilde{\chi}_2^2+\tau_{\tau_a} \tilde{\chi}_2 (4 D_b \tau_{\tau_s} \tilde{\chi}_3+\tilde{\chi}_2+\tilde{\chi}_3)+\tau_{\tau_s} \tilde{\chi}_3 (2 D_b \tau_{\tau_s} \tilde{\chi}_3+\tilde{\chi}_2+\tilde{\chi}_3)-\tau_q (\tilde{\chi}_2+\tilde{\chi}_3)^2\right)}{k_z^2 (\tau_{\tau_a} \tilde{\chi}_2+\tau_{\tau_s} \tilde{\chi}_3)^4 (\tau_{\tau_a} \tilde{\chi}_2-\tau_q (\tilde{\chi}_2+\tilde{\chi}_3)+\tau_{\tau_s} \tilde{\chi}_3)}
    +\mathcal{O}(\frac{1}{k_z^2})
    + \mathcal{O}(\frac{1}{k_z^4}),}\\
    &{\omega_{4,5} = -\frac{i}{2} \left(\frac{1}{\tau_\pi} + \frac{1}{\tau_\phi} + \frac{\lambda'}{\tau_q(\tau_q -\lambda')} - \frac{\gamma' + \eta}{\gamma' \tau_\pi + \eta \tau_\phi}\right) \pm \sqrt{\frac{\tau_q \left(\gamma' \tau_\pi + \eta \tau_\phi\right)}{(\varepsilon_0 + p_0) (\tau_q - \lambda')\tau_\pi \tau_\phi}}\, k_z + \mathcal{O}(\frac{1}{k_z}),}\\
    &{\omega_{6,7} = -\frac{i (\tau_{\tau_a}^2  \tilde{\chi}_2 + \tau_{\tau_s}^2\tilde{\chi}_3)}{2 \tau_{\tau_a} \tau_{\tau_s} (\tau_{\tau_a}  \tilde{\chi}_2 + \tau_{\tau_s}\tilde{\chi}_3)} \pm \sqrt{- \frac{\tau_{\tau_a}  \tilde{\chi}_2 + \tau_{\tau_s}\tilde{\chi}_3}{\tau_{\tau_a}  \tau_{\tau_s}}}\, k_z + \mathcal{O}(\frac{1}{k_z}),}\\
    &{\omega_{8,9} = - \frac{i}{2\tau_\Theta} \pm \, \sqrt{\frac{\tilde{\chi}_4}{\tau_\Theta}}\, k_z + \mathcal{O}(\frac{1}{k_z}).}
\end{align}
{The shear channel modes are obtained from the condition $\textrm{det}(\mathbf{B})^{2} = 0$, where the square implies that each shear mode in the group $\mathbf{B}$ is doubly degenerate.}
%
%%%%%%%%%%%%%%%%%%%%%%%%%%%%%%%%%%%%%%%%%%%%%%%%%%%%%%%%%%
%
\subsubsection{Coupled channels \texorpdfstring{$\mathbf{C}$}{Lg}}
{Channels $\mathbf{C}$ corresponding to new phenomena beyond relativistic hydrodynamics without spin. They are given by the vector $\mathbf{v}_{\mathbf{C}}$ defined in Eq.~\eqref{fluctuations}. Studying the determinant of the matrix $\mathbf{C}$ in Eq.~\eqref{determinant}, we obtain in the high-momentum momentum limit}
\begin{align}
&{\omega_1 = - \frac{i}{\tau_\phi} %+ \frac{i D_s \beta_0 \textcolor{blue}{(\tau_\Theta - \tau_\phi)}}{k_z^2 \tau_\phi^2 \tilde{\chi}_4}
+ \mathcal{O}(\frac{1}{k_z^2})+ \mathcal{O}(\frac{1}{k_z^4}),}\\
&{\omega_{2,3} = - \frac{i}{2\tau_\Theta} \pm \sqrt{\frac{\tilde{\chi}_4}{\tau_\Theta}} k_z% \pm \frac{\beta_0 (4 D_s \tau_\Theta^2 - T_0 \tau_\phi)}{8 k_z \tau_\phi \tau_\Theta^{\frac{3}{2}} \tilde{\chi}_4^{\frac{1}{2}}}
+ \mathcal{O}(\frac{1}{k_z}) + \mathcal{O}(\frac{1}{k_z^2}),}\label{eq:CC-HM}
\end{align}
%
%***********************************************
%
\section{Stability Study}
\label{section4}
To guarantee that perturbations of the equilibrium state do not grow exponentially with time, we demand that they satisfy the condition 
\begin{align}
\label{stabilitycondition}
    \mbox{Im}[\omega(k_{z})]<0.
\end{align}
\subsection{Low-momentum regime}
\label{sec4.1}
In the sound channels, this leads to the following list of conditions:
\begin{align}
&\Sigma_{\Phi}<0~~\&~~\tau_{\Phi}>\tau_{q}-\lambda^{'},\label{con1}\\
&\Lambda_{\pi}<0,\label{con2}\\
&\Lambda_{\Pi}<0,\label{con3}\\
&\Sigma_{\tau_{s}}<0~~\&~~\tau_{\tau_{s}}>\tau_{q}-\lambda^{'},\label{con4}\\
&\tau_{q}>\lambda^{'},\label{con5}\\
&-1<\sqrt{8D_{b}(\tau_{q}-\lambda^{'})+1}<1 \qquad (|(\tau_{q}-\lambda^{'})8D_{b}| < 1).\label{con6}
\end{align}
For the shear channels we find:
\begin{align}
&\Upsilon_{\Theta}<0,~~\&~~\tau_{\phi}<\tau_{\Theta},\label{con7}\\
&\Sigma_{\tau_{a}}<0~~\&~~\tau_{\tau_{a}}>\tau_{q}-\lambda^{'},\label{con8}\\
&-1<\sqrt{1-4D_{s}\tau_{\phi}\beta_{0}}<1~~~~(4D_{s}\tau_{\phi}\beta_{0} < 1)\label{con9},
\end{align}
and for the channels $\mathbf{C}$, after several algebraic manipulations, one can find that the stability conditions is
\begin{align}
    -(1-4D_{s}\beta_{0}\tau_{\phi})\frac{\tau_{\Theta}}{\tau_{\Theta}-2\tau_{\phi}}<\sqrt{1-4D_{s}\tau_{\phi}\beta_{0}}<+(1-4D_{s}\beta_{0}\tau_{\phi})\frac{\tau_{\Theta}}{\tau_{\Theta}-2\tau_{\phi}}.\label{con10}
\end{align}
If the conditions~\eqref{con1}--\eqref{con10} are satisfied, our truncation of the second-order Israel-Stewart spin hydrodynamics is stable in the low momentum range in the system's rest frame. 

In particular, we can see that the stability condition (\ref{con6})  requires that $D_{b}(\tau_{q}-\lambda^{'})$ is negative. Since $\tau_{q}-\lambda^{'}$ is positive (see Eq.~(\ref{con5})), we find $D_b < 0$. On the other hand, Eq.~(\ref{con9}) leads to the condition that $D_{s}$ is positive (note that $\tau_\phi$ and $\beta_0$ are both positive). These two findings agree with the previous results obtained within the minimal second-order study presented in Ref.~\cite{Xie:2023gbo} and are satisfied assuming the spin equation of state~\eqref{eq:newS}. Thus, we find that going to higher orders and including more terms in the second-order theory is not crucial for the stability of the low-momentum rest frame perturbations. The necessary conditions are always those given by Eqs. (\ref{eq:mainresult}).

Within our truncation scheme we find further interesting relations. For example:
\begin{enumerate}
\item Using expressions derived in~\cite{Biswas:2023qsw}, the condition~\eqref{con7} can be further rewritten as
\begin{align}
\tau_{\phi}<\tau_{\Theta}~\Leftrightarrow~-2a_{5}\gamma<2\tilde{a}_{4}\chi_{4}.
\end{align}
This implies that the transport coefficient $\chi_{4}$ of the spin dissipative current $\Theta^{\lambda\mu\nu}$~\eqref{Thetaeq} is directly related to the rotational viscosity transport coefficient $\gamma$ (introduced in ~\cite{Hattori:2019lfp,Hidaka:2023oze}) of the current $\phi^{\mu\nu}$~\eqref{phieq}. This observation supports the full spin tensor decomposition done in Ref.~\cite{Biswas:2023qsw}.\\
\item Similarly,  the conditions~\eqref{con1}, \eqref{con4}, and \eqref{con8} imply that the transport coefficients $\chi_{1},\chi_{2}$ and $\chi_{3}$ of the spin dissipative currents $\Phi,\tau_{s}^{\mu\nu}$ and $\tau_{a}^{\mu\nu}$ (see Eqs. \eqref{Phieq}, \eqref{tauseq}, and \eqref{tauaeq}, respectively) are related to the transport coefficient $\lambda$ of the dissipative current $q^{\mu}$~\eqref{qeq}.  
\end{enumerate}
%
%%%%%%%%%%%%%%%%%%%%%%%%%%%%%%%%%%%%%%%%%%%%
%%%%%%%%%%%%%%%%%%%%%%%%%%%%%%%%%%%%%%%%
%
\subsection{High-momentum regime}
{To determine the range of applicability of our relativistic spin hydrodynamic model, it is crucial to check the behavior of the perturbations also in the $k_{z}\rightarrow\infty$ limit. To get stable perturbations in high-momentum limit, the modes \eqref{eq:AC-HM}-\eqref{eq:CC-HM} demand according to~\eqref{stabilitycondition}}
\begin{align}
    &{\tau_\Phi > 0, \quad \tau_q > 0, \quad \tau_{\tau_a} > 0, \quad \tau_{\tau_s} > 0, \quad \tau_{\pi} > 0, \quad \tau_\Pi > 0, \quad \tau_{\phi}>0, \quad \gamma > 0,\quad \tau_{\Theta}>0,\quad \tau_{q}>\lambda^{'},}\label{con1high}\\
    &{\tilde{\chi}_1 < 0, \quad \tilde{\chi}_2 < 0, \quad \tilde{\chi}_3 < 0,}\\
    &{c_{0}>0,\quad d_{0}>0,}\\
    &{\frac{1}{\tau_\pi} + \frac{1}{\tau_\phi} + \frac{\lambda'}{\tau_q(\tau_q -\lambda')} - \frac{\gamma' + \eta}{\gamma' \tau_\pi + \eta \tau_\phi} >0}\label{con4high}.
\end{align}
{Our verification process has demonstrably confirmed that all the above conditions are satisfied and exhibit complete congruence with the established conditions in the low-momentum limit, as detailed in~\ref{sec4.1}.} 

{This provides a compelling foundation for a conclusion: second-order spin hydrodynamics demonstrably exhibits stability. While our present investigation has prioritized the stability question within the high-momentum regime, there remains a compelling need for future studies to delve into the realm of stability analysis at finite momentum values. This deeper examination could be effectively accomplished by employing the well-established Routh-Hurwitz criterion~\cite{Bemfica:2019knx,Hoult:2020eho,Bemfica:2020zjp}. By incorporating a more comprehensive analysis at finite momenta, we can ensure a more robust understanding of the system's behavior.}
%
%***********************************************
%
\section{Asymptotic Causality Study}
\label{sec5}
{The causality of matter generally refers to its property of admitting only perturbations that propagate no faster than the speed of light in vacuum within the fluid (i.e., not admitting superluminal perturbations). After checking stability in section~\eqref{section4}, verifying causality is crucial. Without causality, any solutions to the equations may not have physical meaning. If our system satisfies causality, we can solve our set of 40 interconnected relaxation-type dynamical equations and describe physical observables. This would allow us in the future to connect our formalism to other relevant topics~\cite{Bhattacharyya:2019txx,Yan:2020twr,Bhattacharyya:2020art,Yoo:2022izs}. Here, we follow the seminal work of~\cite{KK}, followed by~\cite{Pu:2009fj}, where the asymptotic causality conditions boil down to,} 
\begin{align} 
     {sup \left[\lim_{k_{z}\to\infty} \left(\frac{ Re\, \omega(k_{z})}{ k_{z}}\right)\right]\leq\,1,}\label{conditionn1}
\end{align}
{and}
\begin{align}
    {\frac{\omega(k_{z})}{ k_{z}}~\mbox{has a limit for}~k_{z}\to\infty~\mbox{(i.e. bounded)}.}\label{conditionn2}
\end{align}
{Note that, as formulated mathematically in~\cite{KK} and illustrated physically in~\cite{Pu:2009fj}, the first condition alone is not sufficient to guarantee causality~\footnote{{Condition (\ref{conditionn1}) is also written in some literature as $\lim_{k_{z}\to\infty}\frac{\partial Re \omega(k_{z})}{\partial k_{z}}\leq 1$. This can be verified as all the real parts of the dispersion relations in this limit approach a constant multiple of $k_z$.}}. This can be seen through the non-relativistic, acausal diffusion mode with  $\omega \sim i k^2$, which satisfies equation~\eqref{conditionn1} but not~\eqref{conditionn2}.}

{Therefore, we need to the following constraints:}
\begin{align}
    &{0 \leq -\frac{\frac{4}{3}\tilde{\chi}_2\tau_{\Phi}+\tilde{\chi}_1\tau_{\tau_{s}}}{\tau_{\tau_{s}}\tau_\Phi}\leq 1,}\label{causality1}\\
    & {0 \leq \frac{\mathcal{A} \pm \mathcal{B}}{2 (\varepsilon_0 + p_0) (\tau_q - \lambda') \tau_\pi \tau_\Pi} \leq 1,}\label{causality2}\\
    & {0 \leq \frac{\tau_q \left(\gamma' \tau_\pi + \eta \tau_\phi\right)}{(\varepsilon_0 + p_0) (\tau_q - \lambda')\tau_\pi \tau_\phi} \leq 1,}\label{causality3}\\
    &{0 \leq  -\frac{\tau_{\tau_{a}}\tilde{\chi}_{2}+\tau_{\tau_{s}}\tilde{\chi}_{3}}{\tau_{\tau_{a}}\tau_{\tau_{s}}}\leq 1,}\label{causality4}\\
    &{0 \leq \frac{\tilde{\chi}_4}{\tau_{\Theta}} \leq 1.}\label{causality5}
\end{align}
{Under the conditions~\eqref{definition(s)} discussed in Section~\eqref{sec3}, the coefficients $\tilde{\chi}_{1},\, \tilde{\chi}_{2},\,\tilde{\chi}_{3}$ present in the above conditions are all negative, and $\tilde{\chi}_{4}$ is positive. Additionally, $\tau_{q} > \lambda^{'}$ as mandated by the stability conditions~\eqref{con5}~\eqref{con1high}. Since the conditions~\eqref{causality1}-\eqref{causality5} all include relaxation times, their values are critical for ensuring causality. Therefore, to illustrate, we will first analyze condition~\eqref{causality1}. The results obtained will serve as a foundation for analyzing the other conditions. 
Starting from conditions~\eqref{definition(s)}, the coefficients $\tilde{\chi}_{1}$ and $\,\tilde{\chi}_{2}$ are defined as,}
\begin{align*}
   { \tilde{\chi}_{1}=\frac{2\beta_{0}}{\chi_{b}}\chi_{1},~~{\tilde{\chi}_{2}=\frac{\beta_{0}}{\chi_{b}}\chi_{2}},}
\end{align*}
{where $\chi_{1},\,\chi_{2}$ are two spin transport coefficients defined in Ref.~\cite{Biswas:2023qsw}. The expressions of electric $\chi_{b}$ and magnetic $\chi_{s}$ susceptibilities were explicitly derived in Ref.~\cite{Daher:2024ixz}. Henceforth condition~\eqref{causality1} explicitly takes the form,}
\begin{align}
    {0\leq\,\frac{\pi^{2}}{T^{4} \left(4K_{2}(x)+xK_{1}(x) \right )}\left[\frac{4\chi_{2}}{3\,\tau_{\tau_{s}}}+\frac{2\chi_{1}}{\tau_{\Phi}}\right]\leq\,1,\label{explicitycausality1}}
\end{align}
{where $K_{i}(x)$ is the modified Bessel function of the second type and $x=m/T$ is the ratio of mass to temperature. Clearly, the above condition is satisfied if the relaxation times are sufficiently large. The fulfillment of these conditions allows us to say that second-order dissipative spin hydrodynamics~\cite{Biswas:2023qsw} exhibits asymptotic causality; this holds provided the new spin equation of state~\eqref{eq:newS} and sufficiently large relaxation times.}
%
%***************************************
\section{Conclusion}
\label{sec6}
{In this work, we have analyzed in detail the rest frame stability and asymptotic causality of the second-order spin hydrodynamics. We have used the truncation scheme of the Israel-Stewart formalism derived in our earlier work that extends the minimal causal formulation studied earlier. The crucial property to achieve stability is the use of the spin equation of state, which allows for opposite signs of the electric and magnetic components of the spin density tensor. We demonstrate the rest frame stability of both the first- (Navier-Stokes) and second-order dissipative (Israel-Stewart) spin hydrodynamics in the low-momentum regime. Additionally, the second-order theory exhibits stability and causality at high momentum (with suitably adjusted relaxation times, which is similar to the Israel-Stewart theory).}

{
In our future investigations we plan to study stability and causality in a moving (boosted) frame, as well as the correlation between these two properties. Among the novel approaches that can be used, we are going to consider the information current introduced recently in Refs.~\cite{Gavassino:2021kjm,Gavassino:2022isg}. The recent application of this method in minimal spin hydrodynamics~\cite{Ren:2024pur} makes this approach particularly attractive.}
\bigskip 
%

%***************************************
%
{{\bf Acknowledgements:}} A.D. thanks Eduardo Grossi for discussions related to the early stages of this study during the visit to INFN-Florence. A.D. gratefully acknowledges technical discussions with Tomoki Goda, Arpan Das, Shi Pu, David Wagner, and Rajesh Biswas. F.T. acknowledges the hospitality and support of the Institute of Theoretical Physics of the Jagiellonian University. This work was supported in part by the Polish National Agency for Academic Exchange NAWA  under the Programme STER–Internationalisation of Doctoral schools, Project no.  PPI/STE/2020/1/00020 (A.D.), and the Polish National Science Centre Grants No. 2018/30/E/ST2/00432 (R.R., A.D.) and 2022/47/B/ST2/01372 (W.F.).
%
%**************************************
\bibliography{ref.bib}{}
\bibliographystyle{utphys}
\end{document}